\documentclass[conference]{IEEEtran}
\IEEEoverridecommandlockouts
\usepackage{cite}
\usepackage{amsmath,amssymb,amsfonts}
\usepackage{algorithmic}
\usepackage{graphicx}
\usepackage{booktabs}
\usepackage{array}
\usepackage{graphicx} 
\usepackage{float}
\usepackage{comment}
\usepackage{xurl}                     
\usepackage[breaklinks]{hyperref}    
\usepackage{mathtools}               
\allowdisplaybreaks                   
\usepackage{adjustbox}                
\usepackage{graphicx}                 
\usepackage{longtable}                
\usepackage{url} 
\usepackage{caption}  
\usepackage{textcomp}
\usepackage{xcolor}
\usepackage[table]{xcolor} 
\usepackage{booktabs}     
\usepackage{multirow}      
\usepackage{caption}     
\usepackage{booktabs}
\usepackage{multirow}

\usepackage{tikz}
\usetikzlibrary{arrows.meta,shapes,positioning} 

\def\BibTeX{{\rm B\kern-.05em{\sc i\kern-.025em b}\kern-.08em
    T\kern-.1667em\lower.7ex\hbox{E}\kern-.125emX}}
\begin{document}

\title{Quantum Annealing for Staff Scheduling in Educational Environments\\
}

\author{
\IEEEauthorblockN{Alessia Ciacco\IEEEauthorrefmark{1},
Francesca Guerriero\IEEEauthorrefmark{2},
Eneko Osaba\IEEEauthorrefmark{3}}

\IEEEauthorblockA{\IEEEauthorrefmark{1}Department of Political and Social Sciences,
University of Calabria, Rende (CS), Italy}

\IEEEauthorblockA{\IEEEauthorrefmark{2}Department of Mechanical, Energy and Management Engineering,
University of Calabria, Rende (CS), Italy}

\IEEEauthorblockA{\IEEEauthorrefmark{3}TECNALIA, Basque Research and Technology Alliance (BRTA),
Derio (Bizkaia), Spain}

\IEEEauthorblockA{Emails: alessia.ciacco@unical.it,
francesca.guerriero@unical.it,
eneko.osaba@tecnalia.com}
}

\maketitle

\begin{abstract}
We address a novel staff allocation problem that arises in the organization of collaborators among multiple school sites and educational levels. The problem emerges from a real case study in a public school in Calabria, Italy, where staff members must be distributed across kindergartens, primary, and secondary schools under constraints of availability, competencies, and fairness. To tackle this problem, we develop an optimization model and investigate a solution approach based on quantum annealing. Our computational experiments on real-world data show that quantum annealing is capable of producing balanced assignments in short runtimes. These results provide evidence of the practical applicability of quantum optimization methods in educational scheduling and, more broadly, in complex resource allocation tasks.
\end{abstract}

\section{Introduction}
In recent years, the Italian school system has experienced a significant increase in the complexity of its organizational processes. Schools now operate in a highly regulated environment, characterized by increasingly stringent legal constraints deriving from both national legislation and regional directives, alongside a persistent emphasis on cost efficiency and service quality. These structural conditions are further compounded by growing requirements related to safety, inclusion, and accessibility, the need to manage unpredictable staff absences, fluctuations in student numbers, and the increasing diversification of educational and organizational needs. In the Italian context, these factors are particularly relevant, as public schools provide a universal service subject to centrally defined budgetary, staffing, and scheduling constraints, while being locally managed with limited operational flexibility.
Recent regulatory developments have further intensified this complexity. In particular, a recent reform has promoted the consolidation of campuses and services, leading to a significant increase in the average size of educational institutions. This new institutional configuration has made the allocation of personnel across shifts and locations more demanding, as it requires managing a larger number of constraints, spaces, and schedules within an already highly structured regulatory framework. As a result, the day-to-day management of schools increasingly requires more sophisticated planning and decision-support tools than those traditionally employed.
In this scenario, the role of non-teaching school staff, and in particular that of school collaborators, a key figure in the Italian system, has gained strategic importance. Far from serving as mere logistical support, these personnel represent a critical operational resource: their presence and spatial-temporal distribution within the institution, as well as compliance with national contractual regulations, directly affect the regularity of school activities, the cleanliness and maintenance of facilities, student safety and assistance, the supervision of common areas, and emergency management.
Furthermore, in Italy, school collaborators also perform administrative support tasks and provide assistance to students with disabilities. This further underscores the need for careful and well-structured planning that fully respects the existing regulatory framework and established organizational best practices. Consequently, the effective organization of this role has a direct impact on the quality of the educational experience for students and teaching staff, as well as on schools’ ability to comply with national standards and safety requirements.

Formally, the assignment of school collaborators to shifts and operational areas can be framed as a staff scheduling problem \cite{ernst2004staff, caprara2003models, blochliger2004modeling}, that is, the allocation of personnel to time- and space-constrained activities under a set of operational constraints. The staff scheduling literature is well established and has been extensively developed in a wide range of industrial and service contexts. Applications include nurse scheduling in hospitals \cite{sitompul1990nurse, bradley1991continuous}, technician rostering in industrial environments \cite{bailey1985integrated, brunner2010literature, brucker2011personnel}, and workforce management in call centers \cite{ORMECI201441, alfares2007operator}. To the best of our knowledge, no existing study has yet proposed a model specifically designed for school collaborators, taking into account the organizational, spatial, and regulatory peculiarities of this context. 
As shown in \cite{caprara2003models}, personnel scheduling problems, especially when subject to multiple and complex constraints, belong to the class of NP-hard problems. Consequently, the computational effort required to obtain optimal solutions grows exponentially with the size of the instance, and no algorithms are currently known that can solve large-scale instances of these problems in polynomial time. As a result, exact solution methods become impractical for large real-world applications.

For this reason, the search for alternative computational paradigms and solution strategies represents an ongoing and active line of research. Among the approaches currently being explored, quantum computing has emerged as a potential complementary direction. Although significant theoretical and practical challenges remain, and no clear computational advantage has yet been demonstrated for large-scale real-world instances, quantum-based methods are widely regarded as promising candidates for addressing complex combinatorial optimization problems in the long term.

Within quantum computing, two main paradigms can be identified: gate-based quantum computing and quantum annealing (QA). This work focuses on QA, which has been specifically designed to tackle combinatorial optimization problems \cite{rajak2023quantum, hauke2020perspectives}. By exploiting quantum phenomena such as superposition and tunneling, QA aims to explore large solution spaces more effectively. However, its practical benefits over classical approaches remain the subject of active investigation.

QA has been explored in a wide range of combinatorial optimization problems, including routing~\cite{holliday2025advanced, ciacco2025steiner}, facility location~\cite{ciacco2026facility, malviya2023logistics}, packing~\cite{de2022hybrid, garcia2022comparative}, and personnel scheduling. In particular, existing studies have explored its application to nurse scheduling \cite{ikeda2019application}, call center shift scheduling \cite{li2023quantum}, and industrial rostering \cite{hamada2022practical}, highlighting both its potential and its current limitations.
A comprehensive overview of quantum approaches to scheduling and related optimization problems can be found in the survey by \cite{ciacco2025review}.

The application of quantum annealing is explored to evaluate the effectiveness of the proposed staff scheduling model using D-Wave’s LeapCQMHybrid solver. The analysis focuses on the practical applicability of a hybrid quantum-inspired optimization approach, assessed in terms of solution quality, robustness, and scalability on realistic problem instances. The proposed scheduling problem is formulated within D-Wave’s Constrained Quadratic Model (CQM) framework, which combines classical heuristic techniques with quantum-guided exploration and allows the integration of linear constraints together with binary and integer decision variables within a unified optimization framework. This approach enables practical experimentation on realistic instances even in non-industrial contexts, opening new perspectives for structured and data-driven resource management in public schools.

\begin{figure*}[h!]
    \centering
\includegraphics[width=0.8\linewidth]{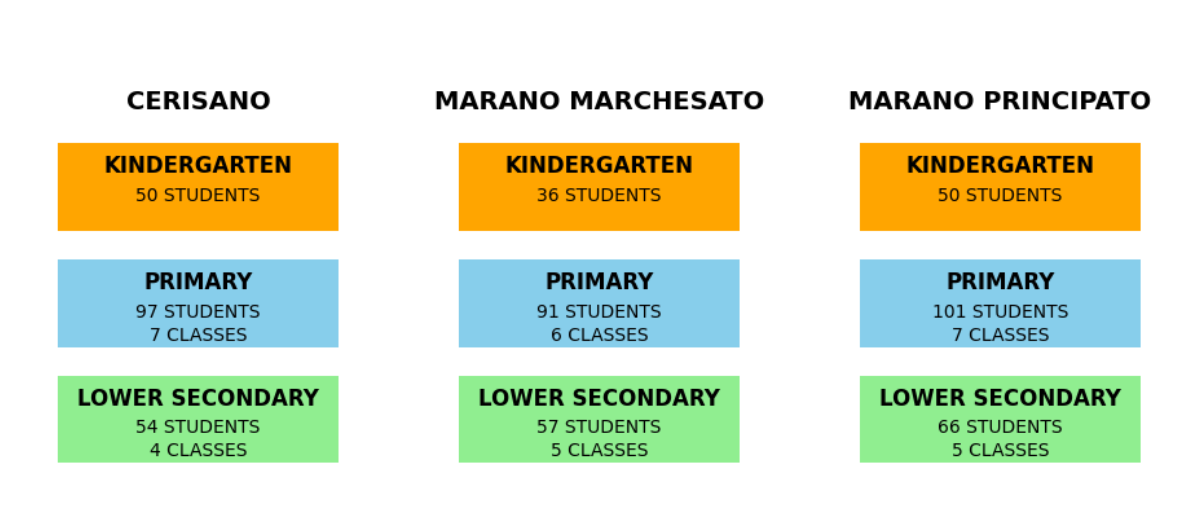}
    \caption{Distribution of students and classes across the three campuses of the Istituto Comprensivo di Cerisano (Cerisano, Marano Marchesato, and Marano Principato), broken down by school level: kindergarten, primary school, and lower secondary school, with the corresponding number of students and classes.
}  \label{scuola}
\end{figure*}

This study proposes an original contribution by focusing on the Italian school context, taking as a reference case the Istituto Comprensivo di Cerisano, located in the province of Cosenza. This medium-sized school has a student population of 629 distributed across 42 classes and is organized into several campuses, including kindergartens, primary, and lower secondary schools, located in Cerisano, Marano Marchesato and Marano Principato. In total, the institution manages nine campuses: three kindergartens, three primary schools, and three lower secondary schools. 
Fig. \ref{scuola} shows the organization of the Istituto Comprensivo di Cerisano by school level and campus, including the number of students and classes, which is relevant for defining the problems used in the work.

Managing the 20 available school collaborators presents additional complexity due to the presence of both full-time and part-time staff, as well as specific constraints related to campus assignments and gender composition required for certain services. These factors highlight the need for a sophisticated scheduling model capable of ensuring optimal coverage of shifts and campuses while respecting regulatory, organizational, and equity constraints.
Furthermore, we also propose carefully designed larger instances that realistically reflect actual scenarios, allowing us to evaluate solution quality, scalability, and robustness of the proposed scheduling model.
This work introduces a mathematical formulation of the school collaborators’ assignment problem that accounts for the organizational specificities and real-world constraints of the school context.

The remainder of this paper is organized as follows. Section~\ref{sec:problem_statement} provides a comprehensive description of the problem, highlighting its main features and underlying motivations. Section~\ref{sec:mathematical_formulation} presents the formal mathematical model, including the decision variables, objective function and constraints. 
Section~\ref{sec:experiments} reports the computational experiments, including the generation of test instances and a thorough analysis of the results obtained. Finally, Section~\ref{sec:conclusions} presents the conclusions and outlines directions for future research.
\section{Problem definition}\label{sec:problem_statement}

The staff scheduling problem consists of assigning a set of collaborators, denoted by $C$, to educational sites, denoted by $S$, over a weekly planning horizon. The set of sites $S$ includes kindergartens ($K \subseteq S$), primary schools, and lower secondary schools distributed across three municipalities. The planning horizon covers working days $D$ (Monday to Friday), with each day divided into two shifts $J$ (morning and afternoon). Each site $s \in S$ has specific opening times, which are modeled as shifts of duration $q_{s,d,j}$ (in hours). The set $J^{+}$ denotes the set of active shifts $(s,d,j)$ such that $q_{s,d,j} > 0$.

The workforce is heterogeneous, consisting of full-time collaborators $C^{FT} \subseteq C$ and part-time collaborators $C^{PT} \subseteq C$. Full-time staff have a maximum weekly working time $H^{FT}$, while part-time staff are limited to $H^{PT}$ weekly hours. Each collaborator may be assigned to at most one site per shift, and every valid shift must be covered by at least one collaborator. Daily working hours $W_c^d$ are computed as the sum of assigned shift durations, and weekly hours $W_c^{week}$ are obtained by summing daily hours over the planning horizon. If a collaborator is assigned to more than 7.2 hours in a single day, a mandatory 30-minute break $B$ is required.

Institutional rules are also incorporated. For every kindergarten site $s \in K$, at least one female collaborator must be assigned, where $F \subseteq C$ denotes the set of all female collaborators. Some collaborators are constrained to work exclusively in a fixed subset of sites $S_c^{bind} \subseteq S$, while others express preferences $S_c^{pref} \subseteq S$ for certain sites, which are treated as soft constraints in the optimization process. 
The set $S^{cand}_c$ represents the candidate sites where collaborator $c$ can potentially be assigned. It coincides with the binding sites $S^{bind}_c$ when such restrictions exist, and with the entire set of sites $S$ otherwise.
\begin{equation}
    S_{\mathrm{cand}}(c) \;=\;
\begin{cases}
  S_{\mathrm{bind}}(c), & \text{if } S_{\mathrm{bind}}(c)\neq\varnothing, \\
  S, & \text{otherwise.}
\end{cases}
\end{equation}
The objective of the problem is to minimize conflicts in staff scheduling, balancing multiple criteria that reflect both collaborator satisfaction and institutional priorities. In particular, the model seeks to reduce multi-site assignments (where collaborators are spread across several sites in a week), assignments to non-preferred sites (when collaborators are scheduled at locations outside their declared preferences), and deviations from contractual weekly workloads.

Minimizing multi-site assignments is particularly important because collaborators generally prefer to be assigned to a single site, which reduces commuting, simplifies task management and fosters continuity with students. The objective function integrates these three components into a normalized and weighted formulation, ensuring comparability across different scales and alignment with management priorities.



\section{Mathematical Formulation} \label{sec:mathematical_formulation}
This section presents the mathematical formulation of the proposed optimization model. It is organized into three parts: the first defines the decision variables, the second describes the objective function and its components and the third introduces the set of constraints that ensure the feasibility of the problem.

\subsection{Decision Variables}
The model uses the following decision variables:

\begin{description}
\item[$x_{c,s,d,j}$]
$\begin{cases}
1, & \text{if collaborator $c \in C$ is assigned to site} \\
       & \text{ $s \in S^{cand}_c$ on day $d$ during shift $j$, with } \\
       & \text{ $(s,d,j) \in J^{+}$},\\
0, & \text{otherwise}
\end{cases}$
  \item[$y_{c,s}$] $\begin{cases} 
    1, & \text{if collaborator $c\in C$ works at site $s\in S^{cand}_c$ } \\
       & \text{at least once during the week} \\
    0, & \text{otherwise}
  \end{cases}$

  \item[$b_{c,d}$] $\begin{cases} 
    1, & \text{if collaborator $c$ must take a mandatory} \\
       &  \text{30-minute break on day $d$ (shift $> 7.2$ hours)} \\
    0, & \text{otherwise}
  \end{cases}$

\end{description}




\subsection{Objective function}

The problem under consideration can be formulated as a \textit{multi-objective optimization problem}, since it simultaneously addresses three distinct and potentially conflicting goals: reducing collaborator dispersion across multiple sites, minimizing deviations from contractual weekly workloads, and respecting collaborators' site preferences. 
\begin{equation}
\label{eq:multisite_penalty}
\min \; \underbrace{\frac{\displaystyle \sum_{c \in C} \sum_{s \in S^{cand}_c} y_{c,s}}{|C| \cdot |S|}}_{\text{penalty for working in multiple sites}},
\end{equation}
\begin{equation}
\label{eq:weekly_hours_penalty}
\min \underbrace{
\frac{
\begin{array}{l}
\displaystyle \sum_{c \in C_{FT}} 
\Biggl(H_{FT} - \sum_{d \in D} (\sum_{s \in S^{\text{cand}}_c} \sum_{j \in J} q_{s,d,j} x_{c,s,d,j} - b_{c,d})\Biggr) \\
\hline
\displaystyle + \sum_{c \in C_{PT}} 
\Biggl(H_{PT} - \sum_{d \in D} (\sum_{s \in S^{\text{cand}}_c} \sum_{j \in J} q_{s,d,j} x_{c,s,d,j} - b_{c,d})\Biggr)
\end{array}
}
{|C_{FT}|\cdot H_{FT} + |C_{PT}|\cdot H_{PT}}
}_{\text{penalty for deviation from standard weekly hours}}
\end{equation}

\begin{equation}
\label{eq:preference_penalty}
\min \; \underbrace{\frac{\displaystyle \sum_{c \in C}  \sum_{s \in S \setminus S_{\text{pref}}[c]} \sum_{d \in D} \sum_{j \in J} x_{c,s,d,j}}{|C| \cdot |S| \cdot |D| \cdot |J|}}_{\text{penalty for violating site preferences}}.
\end{equation}
The three objective components are defined as follows.
\begin{enumerate}
    \item \textbf{Penalty for assignments across multiple sites.}  
    The first objective component, defined in~\eqref{eq:multisite_penalty}, aims at limiting the dispersion of collaborators across different sites during the planning horizon.  
    The binary variable $y_{c,s}$ indicates whether collaborator $c \in C$ is assigned to site $s \in S^{\text{cand}}_c$ at least once during the week. The total number of collaborator--site assignments is normalized by the maximum possible value $|C|\cdot|S|$, yielding a dimensionless quantity in $[0,1]$. This term penalizes solutions in which collaborators are distributed across a large number of sites.

    \item \textbf{Penalty for deviation from contractual weekly workloads.}  
    The second objective component, defined in~\eqref{eq:weekly_hours_penalty}, accounts for deviations from the contractual weekly workload assigned to each collaborator.  
    Full-time collaborators $c \in C_{FT}$ are expected to work $H_{FT}$ hours per week, while part-time collaborators $c \in C_{PT}$ are expected to work $H_{PT}$ hours. The assigned workload is computed as the sum of the durations $q_{s,d,j}$ of the assigned shifts, net of daily breaks $b_{c,d}$.  
    The total deviation from the contractual targets is normalized by the maximum total contractual workload, $|C_{FT}|\cdot H_{FT} + |C_{PT}|\cdot H_{PT}$, ensuring that this term lies in the interval $[0,1]$.

    \item \textbf{Penalty for assignments to non-preferred sites.}  
    The third objective component, defined in~\eqref{eq:preference_penalty}, captures the extent to which collaborators are assigned to sites outside their expressed preferences.  
    For each collaborator $c$, assignments to sites $s \in S \setminus S_{\text{pref}}[c]$ are penalized by summing the corresponding binary decision variables $x_{c,s,d,j}$ over all days $d \in D$ and shifts $j \in J$. The resulting quantity is normalized by $|C|\cdot|S|\cdot|D|\cdot|J|$, yielding a value in $[0,1]$.
\end{enumerate}

In order to simplify the formulation and improve computational performance, these objectives are aggregated into a single scalar function through a weighted sum:
\begin{equation*}
\min \;
w_1 \cdot {\frac{\displaystyle \sum_{c \in C} \sum_{s \in S^{cand}_c} y_{c,s}}{|C| \cdot |S|}}    
+ 
\end{equation*}
\begin{equation*}
w_2 \cdot \frac{
\begin{array}{l}
\displaystyle \sum_{c \in C_{FT}} 
\Biggl(H_{FT} - \sum_{d \in D} (\sum_{s \in S^{\text{cand}}_c} \sum_{j \in J} q_{s,d,j} x_{c,s,d,j} - b_{c,d})\Biggr) \\
\hline
\displaystyle + \sum_{c \in C_{PT}} 
\Biggl(H_{PT} - \sum_{d \in D} (\sum_{s \in S^{\text{cand}}_c} \sum_{j \in J} q_{s,d,j} x_{c,s,d,j} - b_{c,d})\Biggr)
\end{array}
}
{|C_{FT}|\cdot H_{FT} + |C_{PT}|\cdot H_{PT}}
 \end{equation*}
 \begin{equation}
+ \: w_3 \cdot \frac{\displaystyle \sum_{c \in C}  \sum_{s \in S \setminus S_{\text{pref}}[c]} \sum_{d \in D} \sum_{j \in J} x_{c,s,d,j}}{|C| \cdot |S| \cdot |D| \cdot |J|}
\end{equation}
where the weights $w_1, w_2, w_3$ are defined as forming a convex combination, that is, $\displaystyle \sum_{i=1}^{3} w_i = 1$.

By normalizing all components, the objective terms are made comparable, so that the weights $w_1, w_2, w_3$ meaningfully encode the priorities defined by the school management. The resulting objective function is minimized to produce a schedule that balances site continuity, compliance with contractual workloads, and consideration of collaborators’ preferences in accordance with operational guidelines.

\subsection{Constraints}
\begin{align}
&\sum_{c \in C} x_{c,s,d,j} \ge 1 
\quad \forall s \in S, \; d \in D, \; j \in J^+
\label{constr:coverage} \\
&\sum_{s \in S^{cand}_c} \sum_{j \in J} x_{c,s,d,j} \leq 1 
\quad \forall c \in C, \; \forall d \in D 
\label{eq:max_one_shift_per_day}\\
&y_{c,s} \le \sum_{d \in D}\sum_{j \in J} x_{c,s,d,j} 
\quad \forall c \in C, \; s \in S^{cand}_c
\label{constr:link_x_y1} \\ 
&\sum_{d \in D}\sum_{j \in J} x_{c,s,d,j} \le |D| \cdot |J| \; y_{c,s} 
\quad \forall c \in C, \; s \in S^{cand}_c
\label{constr:link_x_y2} \\ 
&\sum_{c \in F} y_{c,s} \ge 1 
\quad \forall s \in K, if s \in S^{cand}_c 
\label{constr:kindergarten_female} \\
&\sum_{d \in D} \Bigl( \sum_{s \in S^{cand}_c} \sum_{j \in J} q_{s,d,j} \, x_{c,s,d,j} - b_{c,d} \Bigr)
 \le H_{PT} \quad \forall c \in C_{PT}\label{constr:weekly_limit0}\\
& \sum_{d \in D} \Bigl( \sum_{s \in S^{cand}_c} \sum_{j \in J} q_{s,d,j} \, x_{c,s,d,j} - b_{c,d} \Bigr)
 \le H_{FT} \quad \forall c \in C_{FT}
\label{constr:weekly_limit} \\
&\sum_{\substack{s \in S^{cand}_c, j \in J^+,\\ q_{s,d,j} > 7.2} }x_{c,s,d,j} \le |S| \cdot |J| \, b_{c,d} 
\quad \forall c \in C, \; d \in D
\label{constr:long_shift_break1} \\
&b_{c,d} \le \sum_{\substack{
s \in S^{cand}_c, j \in J^+ \\
q_{s,d,j} > 7.2
}}x_{c,s,d,j} 
\quad \forall c \in C, \; d \in D
\label{constr:long_shift_break2} \\
&x_{c,s,d,j} \in \{0,1\} \quad \forall c \in C, \; s \in S^{cand}_c, \; d \in D, \; j \in J^+
\label{constr:binary_x} \\
&y_{c,s} \in \{0,1\} \quad \forall c \in C, \; s \in S^{cand}_c
\label{constr:binary_y} \\
&b_{c,d} \in \{0,1\} \quad \forall c \in C, \; d \in D
\label{constr:binary_b} \\
\end{align}

Constraints \eqref{constr:coverage} guarantee coverage, requiring that every site–day–shift $(s,d,j)$ is staffed by at least one collaborator whenever the shift demand $q_{s,d,j} > 0$.  
Constraints~\eqref{eq:max_one_shift_per_day} ensure that each collaborator $c \in C$ can be assigned to at most one shift on any given day $d \in D$.
Constraints \eqref{constr:link_x_y1}--\eqref{constr:link_x_y2} link the weekly assignment indicators $y_{c,s}$ to the daily assignment variables $x_{c,s,d,j}$. A site indicator $y_{c,s}$ is activated if the collaborator works at least one shift in that site, and the upper bound ensures consistency between weekly and daily assignments.  
Constraints \eqref{constr:kindergarten_female} ensure, for Kindergarten sites, at least one female collaborator must be assigned.  
Constraints \eqref{constr:weekly_limit0}--\eqref{constr:weekly_limit} limit the weekly workload according to contractual agreements: part-time workers $\le H_{PT}$ and full-time workers $\le H_{FT}$.
Constraints \eqref{constr:long_shift_break1}--\eqref{constr:long_shift_break2} enforce break rules: for shifts longer than 7.2 hours, the break indicator $b_{c,d}$ must be active, and conversely, breaks are only counted if a long shift is assigned.  
Constraints \eqref{constr:binary_x} define the nature of variables.

\section{Computational study} \label{sec:experiments} 
The goal of the computational experiments is to evaluate the performance of D-Wave's CQM hybrid solver on the proposed staff scheduling problem. CQM extends standard quadratic models by allowing linear constraints and integrality conditions, making it suitable for mixed-integer optimization problems such as ours. D-Wave provides a proprietary hybrid solver, \texttt{LeapCQMHybrid}, which combines classical heuristics with quantum-guided exploration to achieve a balance between global search and local refinement. {It is important to note that, due to the proprietary nature of the solver and the tight integration of classical and quantum components, it is not possible to explicitly isolate or quantify the individual contribution of the quantum processing unit within the overall optimization process.} 

All computational experiments are performed on a Windows 10 machine with an Intel processor (4 physical cores, 8 threads) and 15.6~GB of RAM, using Python 3 in Jupyter Notebook. Exact optimization experiments use Gurobi Optimizer 11.0.1.
The CQM formulation is implemented via D-Wave's Ocean SDK (\texttt{dimod}~0.12.18, \texttt{dwave-system}~1.28.0).
Quantum queries are executed on D-Wave's Advantage2\_system1.4 QPU (4596 qubits, Zephyr graph). Solver parameters are left at default for consistency. For further information on this solver and its applications, the reader is referred to \cite{developers2022measuring, benson2023cqm, osaba2024solving,osaba2025d}. 

{The model is evaluated on real-world data obtained from partner schools to assess the impact of flexibility on solution quality, and on larger realistic synthetic instances to examine the scalability, efficiency, and robustness of the proposed approach under more challenging conditions. Accordingly, the remainder of this section is organized into two main parts: the first focuses on real-world test instances, while the second considers realistic synthetic instances. Each part is further divided into two subsections, one describing the instance generation process and the other reporting and discussing the corresponding computational results.}

\subsection{Case Study: real-world scheduling}
\subsubsection*{Generation of instances}
This study proposes an original contribution by focusing on the Italian school context, taking as a reference case the \emph{Istituto Comprensivo di Cerisano}, located in the province of Cosenza (Calabria, Italy)\footnote{The official website of the school is available at: \url{https://www.cerisanoscuole.edu.it/}}.
The real-world instances reflect the organizational structure of the school network. The instance generation process was carefully designed to reproduce actual working conditions, staff characteristics, and institutional requirements. Below we describe the main components of the instance design:
\paragraph{Collaborators} 
we consider a team of $20$ school collaborators, composed of both full-time (36 hours per week) and part-time (18 hours for week) employees. Among them, two collaborators (TA and TG) are part-time, while the remaining eighteen are full-time.
The staff includes both male and female collaborators, as gender balance plays an operational role: specifically, kindergartens require the presence of at least one female collaborator.
The female collaborators are CG, SG, DB, RB, FR, RD, TA, and TG.
The remaining twelve collaborators, AP, BA, BE, BG, CA, CM, CG2, MM, PG, ME, and MS, are male. 
\paragraph{Sites and levels} 
the school network consists of three physical sites (Cerisano, Marano Marchesato and Marano Principato), each one divided into three levels (Kindergarten, Primary, and Secondary), for a total of $|S| = 9$ sites. Each site may operate on multiple days and shifts, with specific daily schedules.
\paragraph{Days and shifts} 
the planning horizon covers the five working days of the week, i.e., $D = \{\text{Mon}, \text{Tue}, \text{Wed}, \text{Thu}, \text{Fri}\}$. For each day, we define two nominal shifts: $T1$ (morning) and $T2$ (afternoon). The duration of each shift is expressed in hours and derived from the actual school timetable. 
The detailed weekly timetable for all school sites and levels, including morning ($T1$) and afternoon ($T2$) shifts, is reported in Table~\ref{tab:shifts}. {Durations $q_{s,d,j}$ are computed by parsing the official school timetable and converting start/end times into decimal hours.}
\begin{table*}[t]
\centering
\caption{Weekly timetable structure for each school site and educational level. 
The table is organized with rows representing morning ($T1$) and afternoon ($T2$) shifts for each school and level. 
Columns indicate the weekdays: \textbf{Mon} = Monday, \textbf{Tue} = Tuesday, \textbf{Wed} = Wednesday, \textbf{Thu} = Thursday, and \textbf{Fri} = Friday. 
Morning shifts ($T1$) are highlighted in a very light blue, and afternoon shifts ($T2$) in a slightly darker light blue.}
\label{tab:shifts}
\scriptsize
\begin{tabular}{llccccc}
\toprule
\textbf{Site} & \textbf{Level} & \textbf{Mon} & \textbf{Tue} & \textbf{Wed} & \textbf{Thu} & \textbf{Fri} \\
\midrule

\multirow{2}{*}{Cerisano} & \multirow{2}{*}{Kindergarten} & 
\cellcolor{blue!5}T1: 07:30--14:42 & \cellcolor{blue!5}T1: 07:30--14:42 & \cellcolor{blue!5}T1: 07:30--14:42 & \cellcolor{blue!5}T1: 07:30--14:42 & \cellcolor{blue!5}T1: 07:30--14:42 \\
& & \cellcolor{blue!15}T2: 09:18--16:30 & \cellcolor{blue!15}T2: 09:18--16:30 & \cellcolor{blue!15}T2: 09:18--16:30 & \cellcolor{blue!15}T2: 09:18--16:30 & \cellcolor{blue!15}T2: 09:18--16:30 \\
\midrule
\multirow{2}{*}{Cerisano} & \multirow{2}{*}{Primary} & 
\cellcolor{blue!5}T1: 07:45--16:30 & \cellcolor{blue!5}T1: 07:45--14:15 & \cellcolor{blue!5}T1: 07:45--14:15 & \cellcolor{blue!5}T1: 07:45--16:30 & \cellcolor{blue!5}T1: 07:45--14:15 \\
& & \cellcolor{blue!15}T2: 07:45--16:30 & \cellcolor{blue!15}T2: 07:45--14:15 & \cellcolor{blue!15}T2: 07:45--14:15 & \cellcolor{blue!15}T2: 07:45--16:30 & \cellcolor{blue!15}T2: 07:45--14:15 \\
\midrule
\multirow{2}{*}{Cerisano} & \multirow{2}{*}{Secondary} & 
\cellcolor{blue!5}T1: 07:30--14:42 & \cellcolor{blue!5}T1: 07:30--14:42 & \cellcolor{blue!5}T1: 07:30--14:42 & \cellcolor{blue!5}T1: 07:30--14:42 & \cellcolor{blue!5}T1: 07:30--14:42 \\
& & \cellcolor{blue!15}T2: 11:18--18:30 & \cellcolor{blue!15}T2: 11:18--18:30 & \cellcolor{blue!15}T2: 10:18--17:30 & \cellcolor{blue!15}T2: 10:18--17:30 & \cellcolor{blue!15}T2: 08:00--15:12 \\
\midrule

\multirow{2}{*}{Marchesato} & \multirow{2}{*}{Kindergarten} & 
\cellcolor{blue!5}T1: 07:30--14:42 & \cellcolor{blue!5}T1: 07:30--14:42 & \cellcolor{blue!5}T1: 07:30--14:42 & \cellcolor{blue!5}T1: 07:30--14:42 & \cellcolor{blue!5}T1: 07:30--14:42 \\
& & \cellcolor{blue!15}T2: 09:18--16:30 & \cellcolor{blue!15}T2: 09:18--16:30 & \cellcolor{blue!15}T2: 09:18--16:30 & \cellcolor{blue!15}T2: 09:18--16:30 & \cellcolor{blue!15}T2: 09:18--16:30 \\
\midrule
\multirow{2}{*}{Marchesato} & \multirow{2}{*}{Primary} & 
\cellcolor{blue!5}T1: 07:45--16:30 & \cellcolor{blue!5}T1: 07:45--14:15 & \cellcolor{blue!5}T1: 07:45--14:15 & \cellcolor{blue!5}T1: 07:45--16:30 & \cellcolor{blue!5}T1: 07:45--14:15 \\
& & \cellcolor{blue!15}T2: 07:45--16:30 & \cellcolor{blue!15}T2: 07:45--14:15 & \cellcolor{blue!15}T2: 07:45--14:15 & \cellcolor{blue!15}T2: 07:45--16:30 & \cellcolor{blue!15}T2: 07:45--14:15 \\
\midrule
\multirow{2}{*}{Marchesato} & \multirow{2}{*}{Secondary} & 
\cellcolor{blue!5}T1: 07:45--14:57 & \cellcolor{blue!5}T1: 07:45--14:57 & \cellcolor{blue!5}T1: 07:45--14:57 & \cellcolor{blue!5}T1: 07:45--14:57 & \cellcolor{blue!5}T1: 07:45--14:57 \\
& & \cellcolor{blue!15}T2: 07:45--14:57 & \cellcolor{blue!15}T2: 07:45--14:57 & \cellcolor{blue!15}T2: 11:18--18:30 & \cellcolor{blue!15}T2: 07:45--14:57 & \cellcolor{blue!15}T2: 11:18--18:30 \\
\midrule

\multirow{2}{*}{Principato} & \multirow{2}{*}{Kindergarten} & 
\cellcolor{blue!5}T1: 07:30--14:42 & \cellcolor{blue!5}T1: 07:30--14:42 & \cellcolor{blue!5}T1: 07:30--14:42 & \cellcolor{blue!5}T1: 07:30--14:42 & \cellcolor{blue!5}T1: 07:30--14:42 \\
& & \cellcolor{blue!15}T2: 09:18--16:30 & \cellcolor{blue!15}T2: 09:18--16:30 & \cellcolor{blue!15}T2: 09:18--16:30 & \cellcolor{blue!15}T2: 09:18--16:30 & \cellcolor{blue!15}T2: 09:18--16:30 \\
\midrule
\multirow{2}{*}{Principato} & \multirow{2}{*}{Primary} & 
\cellcolor{blue!5}T1: 07:45--16:30 & \cellcolor{blue!5}T1: 07:45--14:15 & \cellcolor{blue!5}T1: 07:45--14:15 & \cellcolor{blue!5}T1: 07:45--16:30 & \cellcolor{blue!5}T1: 07:45--14:15 \\
& & \cellcolor{blue!15}T2: 07:45--16:30 & \cellcolor{blue!15}T2: 07:45--14:15 & \cellcolor{blue!15}T2: 07:45--14:15 & \cellcolor{blue!15}T2: 07:45--16:30 & \cellcolor{blue!15}T2: 07:45--14:15 \\
\midrule
\multirow{2}{*}{Principato} & \multirow{2}{*}{Secondary} & 
\cellcolor{blue!5}T1: 07:45--14:57 & \cellcolor{blue!5}T1: 07:45--14:57 & \cellcolor{blue!5}T1: 07:45--14:57 & \cellcolor{blue!5}T1: 07:45--14:57 & \cellcolor{blue!5}T1: 07:45--14:57 \\
& & \cellcolor{blue!15}T2: 11:14--15:00 & \cellcolor{blue!15}T2: 07:45--14:57 & \cellcolor{blue!15}T2: 11:18--18:30 & \cellcolor{blue!15}T2: 11:18--18:30 & \cellcolor{blue!15}T2: 10:18--17:30 \\
\bottomrule
\end{tabular}
\end{table*}

\paragraph{Continuity and preferences} 
to capture staff-specific constraints, we introduce two types of preferences:
\begin{itemize}
    \item \emph{Binding continuity}: certain collaborators are strictly associated with one or more sites;
    \item \emph{Non-binding preferences}: collaborators express preferences for working at certain sites, which are encouraged but not mandatory.
\end{itemize}
Table \ref{tab:combined_binding_blocks} summarizes these constraints for the schools considered in this study.

\begin{table}[t]
\centering
\caption{Collaborators with site assignments. The table is divided into two blocks: \textbf{Binding} collaborators must be assigned to the specified site and level, while \textbf{Non-Binding Preferences} collaborators have encouraged site assignments but they are not mandatory.}
\label{tab:combined_binding_blocks}
\begin{tabular}{>{\centering\arraybackslash}m{1.5cm}>{\centering\arraybackslash}m{3cm}>{\centering\arraybackslash}m{2.5cm}}
\hline
\textbf{Collaborator} & \textbf{Site} & \textbf{Level} \\
\hline
\multicolumn{3}{c}{\textbf{Binding}} \\
\hline
CM  & Cerisano       & Secondary          \\
CG  & Cerisano       & Kindergarten       \\
SG  & Cerisano       & Kindergarten       \\
MM  & Marano Marchesato & Primary         \\
DB  & Marano Marchesato & Primary         \\
CG2 & Marano Marchesato & Secondary       \\
RB  & Marano Marchesato & Kindergarten    \\
PG  & Marano Marchesato & Kindergarten    \\
ME  & Marano Marchesato & Secondary       \\
BE  & Marano Principato & Secondary       \\
RD  & Marano Principato & Primary         \\
FR  & Marano Principato & Kindergarten    \\
\hline
\multicolumn{3}{c}{\textbf{Non-Binding Preferences}} \\
\hline
CA  & Cerisano       & All levels          \\
BG  & Cerisano       & Primary             \\
AP  & Cerisano       & Secondary           \\
\hline
\end{tabular}
\end{table}
\paragraph{Objective function weights}
the set of weight configurations $(w_1, w_2, w_3)$ used in the objective function is selected in accordance 
with the school’s internal guidelines regarding operational priorities and the relative importance of the considered criteria. In the real-world instances, the weights are fixed to $w_1 = 0.5$, $w_2 = 0.3$, and $w_3 = 0.2$, 
reflecting a stronger emphasis on limiting multi-site assignments, followed by adherence to contractual weekly workloads and, finally, the satisfaction of collaborators’ site preferences.

\subsubsection*{Numerical results}
The optimization problem is solved using five independent runs on the \texttt{LeapCQMHybrid} solver. {All runs produce the same solution, which is equal to the optimal solution obtained with Gurobi (use as a reference to verify optimality). This result confirms the ability of the proposed approach to reach the optimal solution.} The value of the objective function for each run is 0.070602. The corresponding execution times are 21.977 s, 15.71 s, 15.066 s, 10.659 s, and 14.426 s. The average execution time across the five runs is approximately 15.16 s.

\begin{figure*}[h!]
     \centering
 \includegraphics[width=0.93\linewidth]{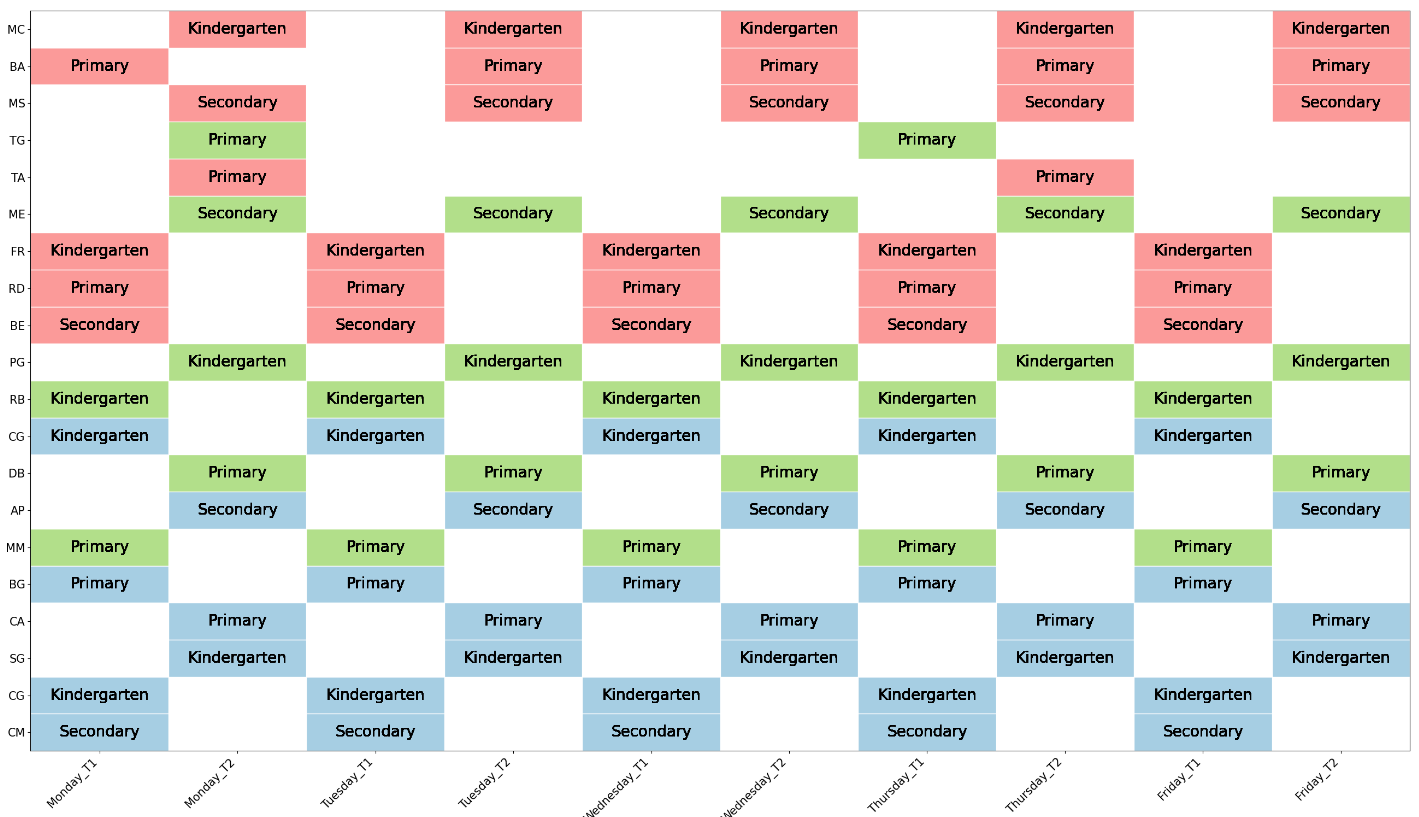}
     \caption{Timetable obtained by the proposed MIP model on the real case study. Rows correspond to collaborators (identified by their initials), while columns represent the weekly time slots (two per day from Monday to Friday). 
 Each colored block indicates the assigned school sites, with distinct colors used to differentiate between institutions. 
 }  \label{fig:schedule}
 \end{figure*}

 Fig.~\ref{fig:schedule} presents the optimized school timetable obtained through our scheduling model. 
 We observe that each class is consistently assigned to its respective school level across the week, with minimal overlaps or empty slots. 
 The color-coded representation highlights how different schools and educational levels are distributed over the time periods, providing an intuitive understanding of the solution structure.
 
 These results indicate that {our approach} not only produces high-quality solutions but also reliably reaches the optimal schedule within a reasonable computational time. The consistency across multiple runs highlights the robustness of the approach and confirms its practical applicability for real-world school timetabling problems.

\subsection{Realistic Large-Scale Scenarios}
\subsubsection*{Generation of instances}
To evaluate the performance and scalability of the proposed scheduling model, we generate larger instances that are proportionally scaled from the real-world case. These synthetic instances mimic the organizational structure and constraints of partner schools, while allowing the assessment of the model on a wider range of problem sizes. The instance generation process is described below.
\paragraph{Collaborators}
we consider a configurable number of collaborators, $n_{\text{collab}}$, including both full-time and part-time staff. To maintain the proportions observed in the real-world instances, $40\%$ of collaborators are female and $60\%$ are male. Moreover, $10\%$ of collaborators are part-time, while the remaining $90\%$ are full-time. This setup ensures that the synthetic instances preserve the workforce composition observed in the reference school network.
\paragraph{Sites and levels}
the number of cities is proportional to the number of collaborators, with approximately $15\%$ of collaborators determining the number of cities ($n_{\text{cities}}$). Each city always contains three educational levels: Kindergarten, Primary, and Secondary, reflecting the typical structure of a comprehensive school in Italy. Therefore, the total number of sites is given by $|S| = n_{\text{cities}} \cdot 3$.
\paragraph{Days and shifts}
the planning horizon covers the standard working days, $D = {\text{Mon}, \text{Tue}, \text{Wed}, \text{Thu}, \text{Fri}}$. Each day includes two nominal shifts: $T1$ (morning) and $T2$ (afternoon). Shift start and end times are based on realistic school schedules and may include slight random variations to introduce diversity. {Shift durations $q_{s,d,j}$ are computed following the same procedure adopted for the real-world instances.} 
\paragraph{Continuity and preferences}
staff-specific constraints are generated proportionally to maintain realistic conditions:
\begin{itemize}
\item {\bf Binding continuity}: approximately $60\%$ of collaborators are strictly assigned to one or more sites, selected randomly;
\item {\bf Non-binding preferences}: approximately $15\%$ of the remaining collaborators express preferred sites within a city. Collaborators already assigned binding sites are excluded from non-binding assignments.
\end{itemize}
\paragraph{Objective function weights}
The objective function weights are set to $w_1 = 0.5$, $w_2 = 0.3$, and $w_3 = 0.2$, consistently with the real-world priorities defined by the school.

\subsubsection*{Numerical results}
We evaluate the performance of the proposed scheduling model on the generated larger instances, varying the number of collaborators from 25 to 40. For each instance size, five independent runs are performed using the \texttt{LeapCQMHybrid} solver, with a time limit of 10 seconds for run. Table~\ref{tab:results_large} reports the obtained results, including the average objective function value, standard deviation of the objective function, average execution time, standard deviation of execution time and the percentage of instances in which the optimal solution found by Gurobi is reached.
\begin{table}[h]
\centering
\caption{Performance of the scheduling model on synthetic large-scale instances. 
Each row reports results for a given number of collaborators $n_{\text{collab}}$, including: 
\textbf{Obj. value (mean)} and \textbf{std. dev.}, 
\textbf{Time (mean, s)} and \textbf{std. dev.}, 
and \textbf{\% optimal} (percentage of runs reaching the Gurobi optimum). 
A value of ``--'' indicates that no valid solution was obtained in any of the five runs.}
\label{tab:results_large}
\setlength{\tabcolsep}{5pt} 
\renewcommand{\arraystretch}{1.2} 
\begin{tabular}{>{\centering\arraybackslash}m{1cm}
                >{\centering\arraybackslash}m{1cm}
                >{\centering\arraybackslash}m{1cm}
                >{\centering\arraybackslash}m{1cm}
                >{\centering\arraybackslash}m{1cm}
                >{\centering\arraybackslash}m{1cm}}
\toprule
$n_{\text{collab}}$ & Obj. value (mean) & Obj. std. dev. & Time (mean, s) & Time std. dev. & \% optimal \\
\midrule
25 & 0.073194 & 0      & 7.9906 & 0.7292 & 100\% \\
30 & 0.055911 & 0      & 7.8598 & 1.3295 & 100\% \\
35 & 0.046204 & 0.00068 & 8.4632 & 0.3486 & 20\% \\
40 & --       & --     & --     & --      & 0\% \\
\bottomrule
\end{tabular}
\end{table}


The results indicate that for smaller instances ($n_{\text{collab}} = 25, 30$), \texttt{LeapCQMHybrid} solver consistently reaches the optimal solution previously obtained by Gurobi. 
As the instance size increases to 35 collaborators, the solver still finds high-quality solutions, but only 20\% of the runs reach the true optimal solution, highlighting the increased difficulty of larger instances. For instances with 40 collaborators, {the solver fails to produce a feasible solution within the imposed time limit as a consequence of the increased problem scale and combinatorial complexity become prohibitively large.}
%
%

Overall, these numerical experiments demonstrate that the proposed model scales effectively for moderately large scenarios, maintaining high solution quality and reasonable computational effort. The analysis also provides insights into the practical limits of current QA approaches for complex scheduling problems, suggesting that hybrid or enhanced techniques may be required for very large instances.
\section{Conclusions}\label{sec:conclusions}
This work addresses an innovative personnel assignment problem in the educational context, focusing on the scheduling of school staff across multiple sites and educational levels. A mathematical optimization model is developed to capture the operational complexities typical of school systems, integrating constraints on staff availability, skills, contractual limits, gender balance, and service continuity. The proposed formulation enables realistic staff allocation while ensuring full coverage of weekly activities and a fair workload distribution.

The model is first applied to a real-world case study, demonstrating its ability to reproduce the existing organizational structure and, in several cases, improve efficiency and balance. The approach is then extended to larger synthetic scenarios to assess scalability and robustness under increasingly complex conditions. Numerical results show that the proposed hybrid quantum-inspired optimization approach consistently produces high-quality and balanced schedules within limited computational times.

Overall, these results confirm the practical applicability and competitiveness of quantum annealing–based hybrid optimization methods for complex combinatorial problems. Their application to the school scheduling context, still relatively underexplored, demonstrates the versatility and potential impact of quantum computing techniques beyond traditional industrial domains, contributing to bridging the gap between theoretical research and real-world applications of quantum optimization in organizational planning.

Future research directions include extending the model to incorporate dynamic constraints and multi-period preferences, integrating predictive components based on historical data and artificial intelligence techniques, and exploring next-generation quantum architectures and advanced hybrid optimization strategies to address larger-scale instances.

\section*{Acknowledgments}
Eneko Osaba acknowledges support from the Basque Government through the ELKARTEK program, project "KUBIBIT - Kuantikaren Berrikuntzarako Ikasketa Teknologikoa" (KK-2025/00079).
\bibliographystyle{apalike}
{\small
\bibliography{rif}}
\end{document}